\documentclass[%
 reprint,superscriptaddress,
 amsmath,amssymb,
 aps,
]{revtex4-2}

\usepackage{dsfont}
\usepackage{tikz}
\usepackage{graphicx}
\newcommand{\Brho}{\boldsymbol{\rho} }

\usepackage{pifont}
\expandafter\let\csname equation*\endcsname\relax
\expandafter\let\csname endequation*\endcsname\relax
\usepackage{graphicx}
\usepackage{amssymb}
\usepackage{amsmath}
\usepackage{amsthm}

\usepackage{hyperref}
\usepackage{epstopdf}
\usepackage{float}
\usepackage{enumitem}
\usepackage{calc}

\usepackage{pdfpages} 
\makeatletter
\AtBeginDocument{\let\LS@rot\@undefined}
\makeatother

\begin{document}

\title{Steady-state selection in multi-species driven diffusive systems}

\author{Luigi Cantini}
\affiliation{  Laboratoire de Physique Théorique et Modélisation (CNRS UMR 8089),
CY Cergy Paris Université, 95302 Cergy-Pontoise, France
}
\author{Ali Zahra}
\affiliation{  Laboratoire de Physique Théorique et Modélisation (CNRS UMR 8089),
CY Cergy Paris Université, 95302 Cergy-Pontoise, France
}
\affiliation{ Departamento de Matem\'atica, Instituto Superior T\'ecnico, Av. Rovisco Pais, 1049-001
Lisboa, Portugal
}

\date{\today}

\begin{abstract}
We introduce a general method to determine the large scale non-equilibrium steady-state properties of one-dimensional multi-species driven diffusive systems with open boundaries, generalizing thus the max-min current principle known for systems with a single type of particles. This method is based on the solution of the Riemann problem of the associated system of conservation laws.
We demonstrate that the effective density of a reservoir depends not only on the corresponding boundary hopping rates but also on the dynamics of the entire system, emphasizing the interplay between bulk and reservoirs. We highlight the role of Riemann variables in establishing
the phase diagram of such systems.
We apply our method to three models of multi-species interacting particle systems and compare the theoretical predictions with numerical simulations. 
\end{abstract}

\maketitle

Driven diffusive systems appear in various areas across 
physics, chemistry, and theoretical biology 
\cite{chou2011non, 
blythe2007nonequilibrium,fang2019nonequilibrium} and are 
widely regarded as a fundamental playground in order to 
understand the behavior of complex systems away from 
thermal equilibrium \cite{schmittmann1995statistical}. 
A classic illustration of such systems involves particles moving within a lattice and subject to 
hard--core exclusion. The introduction of a bias in 
their movement,  simulating the influence of an external driving force, builds up
macroscopic currents in the stationary state. A 
particularly relevant setting consists in putting a 
one-dimensional system in contact with boundary particles 
reservoir, the interplay between boundary dynamics and bulk 
driving leading to genuinely out of equilibrium phenomena 
such as  boundary induced phase transitions 
\cite{krug1991boundary}.
In this case, when the system presents a single species of 
particles, a simple general principle known as the max-min 
current principle \cite{krug1991boundary,
krug1991steady,popkov1999steady,hager2001minimal} allows  
to determine the phase diagram for the steady state current 
and particle density as a function of the boundary 
reservoir densities.
Despite the success of this principle in treating 
one-dimensional open boundary problems,
its generalization to systems containing several different species of particles has been a long-standing challenge \cite{rakos2004exact,popkov2004infinite,bonnin2021two, gupta2023interplay}. 

The goal of the present paper is to put forward a scheme that permits to determine the steady state average particle densities and currents of one-dimensional multi-species driven system with open boundaries. Such a scheme is based essentially on the sole knowledge of the bulk hydrodynamic behavior of the model. 
As a starting point, similarly to the max-min principle, one supposes the boundary densities to be known. In a systems with $n$ different particle species, these are denoted by $\boldsymbol{\rho}^{L}=\{\rho_1^L,\rho_2^L,\dots,\rho_n^L\}$ for the left boundary and $\boldsymbol{\rho}^{R}=\{\rho_1^R,\rho_2^R,\dots,\rho_n^R\}$ for the right boundary. Then the bulk density is determined by the solution of the associated Riemann problem at the origin (RP$_0$)
\begin{equation}\label{principle0}
(\boldsymbol{\rho}^{L},  \boldsymbol{\rho}^{R} ) \xrightarrow[]{\text{RP}_0} \boldsymbol{\rho}^{B}.
\end{equation}
As a first argument in support of this claim,  we shall show  that this principle is equivalent to to Krug's max-min current principle when applied to the case of single-species model. We shall moreover present a further heuristic justification of it based on a vanishing viscosity regularization of the associated conservation laws which applies to general multi-species  case.

By itself the principle (\ref{principle0}) is not enough to 
determine the bulk densities since one has at the same time 
to make sense of the boundary densities. If one supposes 
that the boundary currents are functions of the boundary 
densities alone, then current conservation through the 
entire systems provides the missing conditions to 
completely determine both bulk and boundary densities. 
We apply this scheme to three models, where we 
have access to the particle currents as functions of the 
particle densities (which is necessary in order to solve 
numerically the associated Riemann problem): 2-TASEP with 
arbitrary bulk hopping rates, hierarchical 2-ASEP and 
a 3-TASEP. In all these three model we find good 
agreement with numerical simulations.

\section{The scheme}

The large scale behavior of driven diffusive system consisting of $n$ species of particles  is generally governed by a system of conservation laws
\begin{equation}\label{conservation-1}
\partial_{t}\boldsymbol{\rho} + \partial_{x}\boldsymbol{J} = 0
\end{equation}
where the $n$ locally conserved quantities are the coarse-grained particle densities $ \boldsymbol{\rho}  (x,t) = (\rho_{1}(x,t),...,\rho_{n}(x,t))$, with associated currents
	$\boldsymbol{J}(\boldsymbol{\rho}) = (J_{1}(\boldsymbol{\rho}),.., J_{n}(\boldsymbol{\rho}))$.
When the system is  defined on a finite interval $ x \in [0, L] $ and coupled to two reservoirs with densities $\Brho^L$ and $\Brho^R$ the system reaches in the limit $t \rightarrow \infty$ a steady state 
with uniform bulk densities $\boldsymbol{\rho}^{B}(\Brho^L,\Brho^R)$.
We claim that for $L\rightarrow \infty$, these bulk densities are determined by solving a Riemann problem.
Such a problem is formulated on an infinite line $x \in \mathbb{R}$ with an initial condition consisting of two regions of uniform densities, on the left and on the right of the origin $x=0$
$$\boldsymbol{\rho}  (x,0) =  \boldsymbol{\rho}^{L} \mathds{1}_{x<0}(x) + \boldsymbol{\rho}^{R} \mathds{1}_{x>0}(x) \quad x \in \mathbb{R}.
$$
The solution of the Riemann problem is invariant under the rescaling $(x,t) \rightarrow (\lambda x, \lambda t)$ and therefore takes the form $ \Brho(x,t) = \Brho(\frac{x}{t})$. In particular, for $t>0$, $\boldsymbol{\rho}  (0,t)$ is independent of time, so we define: $\boldsymbol{\rho}|_{0}(\boldsymbol{\rho}^{L},\boldsymbol{\rho}^{R}) := \boldsymbol{\rho}  (0,t)$ and we call it \textit{the solution to the Riemann problem at the origin}.
Our claim is that the bulk densities for the open boundary problem with given boundary densities  coincide with the solution at zero of the corresponding Riemann problem, namely:
\begin{equation}\label{principle}
\boxed{\boldsymbol{\rho}^{B}(\Brho^L,\Brho^R) = \boldsymbol{\rho}|_{0}(\Brho^L,\Brho^R)}
\end{equation}

The exact meaning of the boundary conditions is a 
mathematically subtle issue \cite{bardos1979first, 
dubois1988boundary, mazet1986analyse}. We define them in an 
operative way as the densities of the first and last site 
of the lattice, meaning that the two boundary sites can be 
conceptually considered as part of their nearby reservoirs.
Let us be more specific about the boundary 
dynamics we shall consider. 
At each boundary a particle can either enter or exit the 
system,  or it can change its own species. If we identify 
to empty sites as particles of a species $0$, the dynamics 
is fully encoded in the rates $\boldsymbol{\nu}^{L}=
\{\nu^{L}_{i,j},0\leq i\neq j\leq n\}$ at the left and $
\boldsymbol{\nu}^{R}=\{\nu^{R}_{i,j},0\leq i\neq j\leq n\}$ 
at the right boundary
$$
j \xrightarrow[]{\nu^{L}_{i,j}} i\qquad i \xrightarrow[]
{\nu^{R}_{i,j}} j
$$
The boundary densities $\boldsymbol{\rho}^{L}$ and $
\boldsymbol{\rho}^{R}$, as well as the bulk ones are then 
functions of the boundary rates.

Since the boundary hopping rates are independent of the rest of the system, we can write the current on a given boundary as a function of the density of that boundary only
\begin{equation}\label{equ:boudarycurrents}
    \begin{split}
J_i^{L}(\boldsymbol{\rho}^{L}) = \sum_{j=1}^n \rho_j\nu^L_{ij}  - \rho_i \sum_{j=1}^n \nu^L_{ji}  \\
J_i^{R}(\boldsymbol{\rho}^{R}) = \rho_i \sum_{j=1}^n \nu^R_{ij}  -  \sum_{j=1}^n \rho_j\nu^R_{ji}    \end{split}
\end{equation}
In the steady state, we have
\begin{equation}\label{equ:curr equa}
	\boldsymbol{J}^{L}(\boldsymbol{\rho}^{L}) = \boldsymbol{J}(\boldsymbol{\rho}^{B})= \boldsymbol{J}^{R}(\boldsymbol{\rho}^{R})
	\end{equation}
In conclusion, eqs.(\ref{principle},\ref{equ:curr equa}) provide a system of equation enabling to determine the bulk and boundary densities of the system.
%
%
%


\subsection{Reformulation of the max-minximal Current Principle}

A first argument in favor of the principle eq.(\ref{principle}) is the fact that in the case of a single species of particle it coincides with Krug's max-min current principle.
According to this principle, the steady-state current is obtained as:
\cite{krug1991boundary, popkov1999steady, hager2001minimal, katz_nonequilibrium_1984}
\begin{equation}
\label{eqn:ecp}
j =
\begin{cases}
\max_{\rho \in [\rho^{R}, \rho^{L}]}J(\rho)  & \text{if $\rho^{L}>\rho^{R}$} \\
\min_{\rho \in [\rho^{L}, \rho^{R}]}J(\rho)  & \text{if $\rho^{L}<\rho^{R}$} \\
\end{cases}
\end{equation}
Let's compare this result with what one would obtain by applying eq.(\ref{principle}). Let's start with the case where $\rho^{R} > \rho^{L}$, which corresponds to a minimum current phase. 
When considering the associated Riemann problem we can assume the
current $J$ to be a convex function of the density in the interval $[\rho^{L}, \rho^{R}]$, otherwise one has to replace it  with its  the convex hull in the interval $[\rho^{L}, \rho^{R}]$ \cite{osher1983riemann}.
The solution to the Riemann problem can be expressed as a function of $u = \frac{x}{t}$:
\begin{equation}
\rho  (u) = \rho^{L} {\bf 1}_{u< v(\rho^{L})  } + \rho^{R} {\bf 1}_{u> v(\rho^{R})} +   v^{-1}(u)   {\bf 1}_{ v(\rho^{L}) < u < v(\rho^{R})}
\end{equation}
where $v(\rho) := \frac{dJ}{d\rho} $.
To compare the solution at zero with the density predicted by the minimum current phase, we can identify three cases:\\
1) If $v(\rho^{L}) > 0$, then the solution at zero has a value of $\rho^{L}$, and simultaneously, the minimum $\min_{\rho \in [\rho^{L}, \rho^{R}]}(J(\rho))$ is reached at $\rho^{L}$. In this case, the bulk has the same density as the left boundary, which we refer to as the \textit{left-induced phase}.\\
2) If $v(\rho^{R}) < 0$, then the solution at zero has a value of $\rho^{R}$, and simultaneously, the minimum $\min_{\rho \in [\rho^{L}, \rho^{R}]}(J(\rho))$ is attained at $\rho^{R}$. This is referred to as a \textit{right-induced phase}.\\
3) If neither of the two previous statements is true, there exists, due to the monotonicity of the derivative, a unique value $\rho^{B} \in [\rho^{L}, \rho^{R}]$ for which $v(\rho^{B}) = 0$. This value corresponds to both the Riemann solution at zero and the minimum $\min_{\rho \in [\rho^{L}, \rho^{R}]}(J(\rho))$. We refer to this situation as the \textit{bulk-induced phase}.

When $\rho^{R} < \rho^{L}$, a similar reasoning can be applied, but we replace $J(\rho)$ with its concave hull over the interval $[\rho^{R}, \rho^{L}]$. 
So we conclude that the max-min current principle and the eq.(\ref{principle}) give the same answer.

As an example, in the case of a single-species TASEP, we have $v(\rho^B) = 1-2 \rho^B$. When $v > 0$, we have $\rho^B < \frac{1}{2}$, which corresponds to the low-density phase, and the bulk is left-induced. The high-density regime corresponds to a right-induced bulk density. The maximal current phase, where $\rho = \frac{1}{2}$, is not induced from either the left or the right.

\subsection{Multiple Conserved Quantities}\label{sect:MCQ}

In this section we shall provide a plausibility argument for eq.(\ref{principle}). It will be by no means a proof of that equation, but more support will come from the comparison with simulations, discussed in the next section.
Our argument is based on a vanishing viscosity approach.
This involves adding a diffusive component to the current such that the total current, which remains constant in the steady state, is given by:
\begin{equation}\label{phys-curr}
\boldsymbol{J}^{total} = \boldsymbol{J}(\boldsymbol{\rho}) - \epsilon D(\boldsymbol{\rho}) \frac{\partial \boldsymbol{\rho}}{\partial x}
\end{equation}
Here, $\epsilon > 0$ and $D(\boldsymbol{\rho})$ is a positive-definite matrix. Since the conservation laws become locally scalar in the directions of the eigenvectors of the Jacobian $\frac{\partial J_{i}}{\partial \rho_{j}}$ we assume that this property extends to the viscous case, implying that $D(\boldsymbol{\rho})$ commutes with the Jacobian. This assumption ensures a mathematically stable regularization scheme for the boundary problem.

For the rest of the argument we shall assume that the conservation laws eq.(\ref{conservation-1}) admit $n$ independent Riemann variables  
$\boldsymbol{\xi}=(\xi_1,\dots,\xi_n)$. These are functions of the densities $\boldsymbol{\xi}(\boldsymbol{\rho})$, that "diagonalize" the conservation equations eq.(\ref{conservation-1}), in the sense
$$
\partial_t \xi_i(x,t) + v_i(\boldsymbol{\xi}) \partial_x \xi_i(x,t)=0,
$$
where it can be shown that  the speeds $ v_{k} $ are the eigenvalues  of the Jacobian matrix $\frac{\partial J_{i}}{\partial \rho_{j}}(\boldsymbol{\rho})$. We remark that the existence of the Riemann variables is ensured for $n=1,2$ (for $n=1$ the Riemann variable is the density itself). 
Now, rewriting eq.(\ref{phys-curr}) in terms of the Riemann variables we get the ordinary differential equation:
\begin{equation}\label{key2}
\frac{\partial \boldsymbol{\xi}}{\partial x} = \epsilon^{-1} M^{-1}D^{-1}(J(\boldsymbol{\xi}) - J^{total}) := F(\boldsymbol{\xi})
\end{equation}
where $M_{ij} = \frac{\partial \rho_{i}}{\partial \xi_{j}}$.
In the limit $\epsilon \rightarrow 0$ we have as expected 
$J(\boldsymbol{\xi})= J^{total}$ on all the 
system, with the possible exception of microscopic regions close to the 
boundaries. This means that the bulk value $\boldsymbol{\xi}^B$ represents a stationary point of the ODE (\ref{key2}), $F(\boldsymbol{\xi}^B)=0$.
%
%
%
In order  to determine the relation between the bulk and boundary values of each Riemann variable, we linearize the ODE around the the stationary point. 
It is not difficult to show that the Jacobian matrix $\frac{\partial F}{\partial \boldsymbol{\xi}}$ is diagonal at the stationary point $\frac{\partial F_{i}}{\partial \xi_{j}} (\boldsymbol{\xi}^B) = \epsilon^{-1} d_{i}^{-1} v_{i} \delta_{ij}$, where $d_{i} >0$ are the eigenvalues of the diffusion matrix $D$. 
An illustrative example of the field associated to the ODE for a two-component system is in figure \ref{fig:phase}
%
\begin{itemize}[leftmargin=*]
\item When $v_i < 0$, then $\xi_i(x)$ experiences exponential decay towards the stationary bulk value. The decay rate is given by $\mu_i = \epsilon^{-1} d_{i}^{-1} v_{i}$ . In this scenario, the bulk stationary value is attained on the left side after a boundary layer of typical size $1/ \mu_i$, which is proportional to $\epsilon$. On the right boundary, the system simply extends the bulk behavior, indicating a right-induced phase.
\item When $v_i > 0$, using a similar argument we can infer that $\xi_i$ is induced from the left, and the boundary layer is located on the right.
\item When $v_i =  0$; the size of boundary layer diverges for finite $\epsilon$. The flow of the ODE in the direction of the associated eigenvector indeed ceases to be exponential and becomes rather polynomial. The bulk is therefore not induced by any boundary, however, it belongs to the manifold $v_i(\boldsymbol{\xi}) =  0$. We say that we are in a bulk-induced phase for $\xi_i$.
\end{itemize}
This is the same result one would obtain by considering the solution of Riemann problem at the origin.
Let's point out that the idea of looking at the signs of eigenvalues governing the phase transition in multi-species driven diffusive systems has already been discussed in \cite{popkov2011hierarchy}, however without reference to the Riemann variables.

\section{Application to multi-components interacting particles systems}

In this section we consider three different  driven diffusive systems. The first two contain each two species of particles. More specifically the first one is the $2$--TASEP introduced in \cite{derrida1996statphys,mallick1996shocks}, while the second one is a hierarchical $2$--species ASEP. The third model is a particular case of $3$--species TASEP. 
For all this models we compare numerical simulations 
 with the predictions of the system of equations eq.(\ref{principle}) and eq.(\ref{equ:curr equa}).

This system of equations cannot be solved analytically therefore we make use of an iterative procedure: we 
begin by selecting random initial densities for the 
boundaries. Then, we determine the bulk density using 
equation \ref{principle}, which provides information about 
the current. Subsequently, we calculate the boundary 
densities by inverting equation \ref{equ:boudarycurrents}. 
We continue this iteration process between the boundaries 
and the bulk until convergence is achieved. 
However, it is worth noticing that this algorithm may 
encounter cyclic trajectories. To prevent this issue, we 
introduce a damping parameter $\gamma$, which should be 
chosen sufficiently small. The updated equation becomes:
$\boldsymbol{x}^{n+1} = \gamma \boldsymbol{f}(\boldsymbol{x}^{n}) + (1 - \gamma) \boldsymbol{x}^{n}$ 
Here, $\boldsymbol{x}^n$ represents the set of variables 
after the $n$-th iteration, and $\boldsymbol{f}$ represents 
the set of functions governing the iterations.

\subsection{2-TASEP with arbitrary hopping rates}
	
\begin{figure}[h!]
\centering
\includegraphics[width=.49\linewidth]{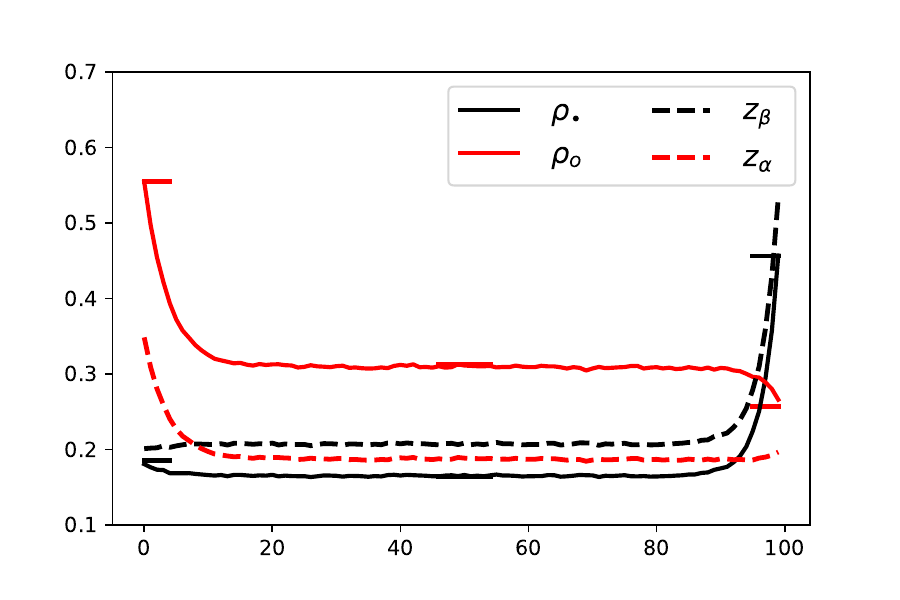}
\includegraphics[width=.49\linewidth]{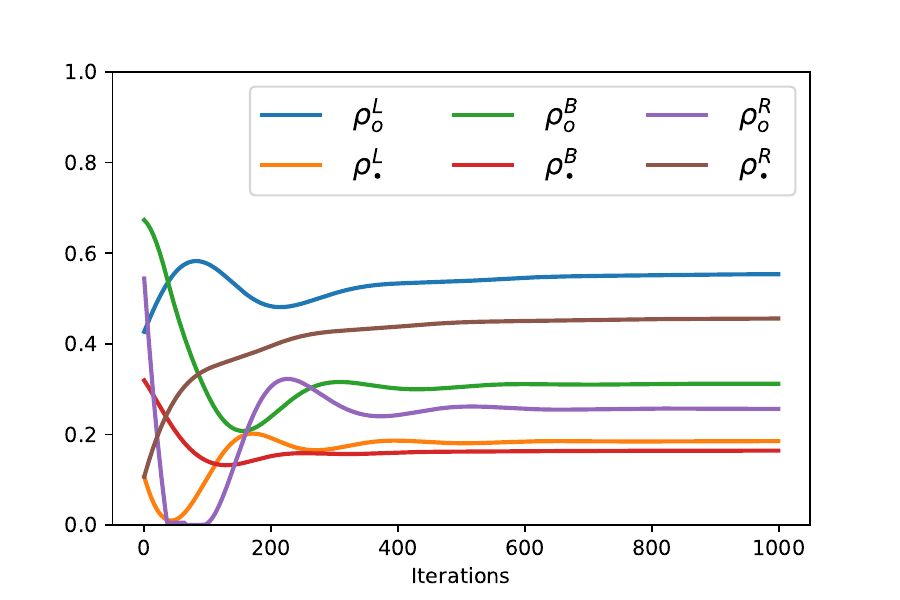}\\
\includegraphics[width=.49\linewidth]{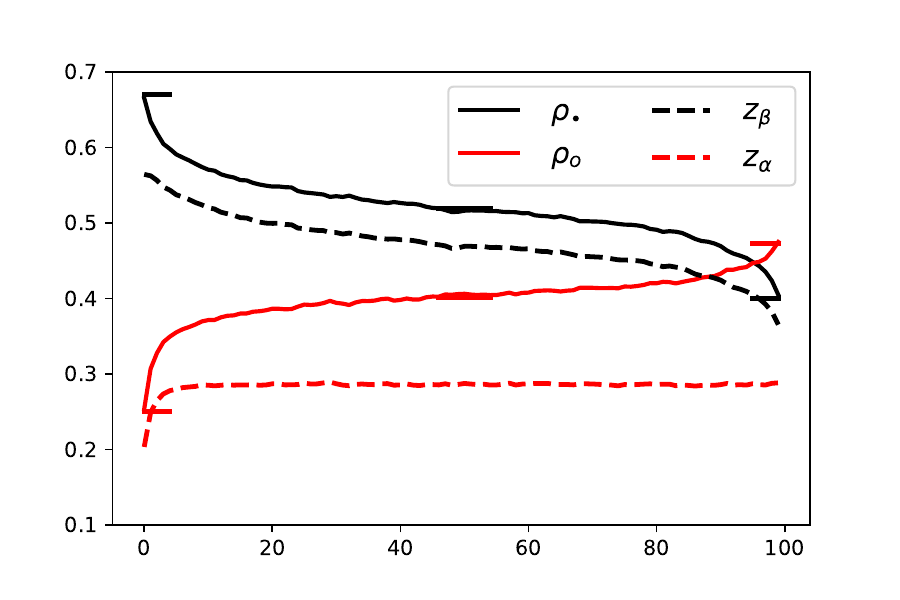}
\includegraphics[width=.49\linewidth]{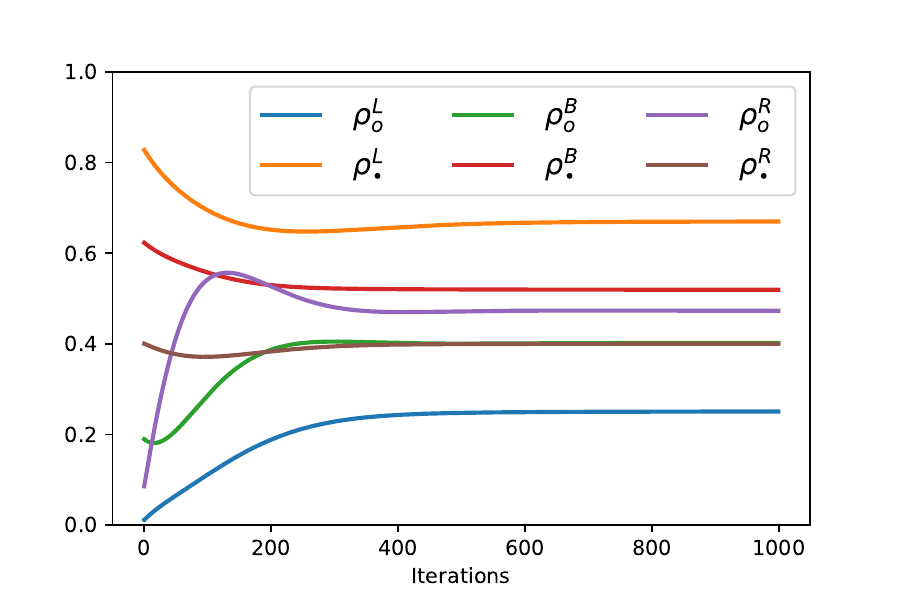}
	
\caption{
On the left, two examples of Monte-Carlo simulation of the density profile for 2-TASEP (continuous lines) along with the corresponding Riemann variables (dashed lines) for a lattice of size $L=100$. The horizontal segments represent the predicted values. On the right, the evolution of densities for the iterative algorithm with damping $\gamma = 0.01$ (up to  1000 iterations). Parameters values for top diagrams: $\alpha=0.5$, $\beta=1.5$, $(\nu_{\bullet \circ}^{R},\nu_{\ast \circ}^{R},\nu_{\bullet \ast}^{R}) = (0.29,0.08,0.07) $, $(\nu_{\bullet \circ}^{L},\nu_{\ast \circ}^{L},\nu_{\bullet \ast}^{L}) = (0.24,0.04,0.12)$. For the bottom diagrams: $\alpha=0.4$, $\beta=0.7$, $(\nu_{\bullet \circ}^{R},\nu_{\ast \circ}^{R},\nu_{\bullet \ast}^{R}) =  (0.5,0.1,0.8)$, $(\nu_{\bullet \circ}^{L},\nu_{\ast \circ}^{L},\nu_{\bullet \ast}^{L}) = (0.1,0.2,0.5)$.}
	\label{fig:2tasep}
\end{figure}

This first model is a two-species generalization of TASEP, it 
consists of two types of particles, denoted by $
\bullet$ and $\circ$,  (empty sites are denoted by 
$\ast$). The hopping rates in the bulk are :
$$
\bullet \ast \xrightarrow[]{\beta} \ast\bullet\qquad
\ast \circ \xrightarrow[]{\alpha} \circ\ast\qquad
\bullet \circ \xrightarrow[]{1} \circ\bullet
$$
while the only non vanishing boundary rates we consider are 
$\nu^{L/R}_{\bullet\ast},\nu^{L/R}_{\ast\circ},\nu^{L/R}_{\bullet\circ}$. 
The currents for this model
have been calculated in \cite{cantini2008algebraic} and used in \cite{cantini2022hydrodynamic} in order to study its hydrodynamic behavior and in particular to solve the corresponding Riemann problem. Let's recall the expression of the currents:
	\begin{gather}
	J_\circ(\rho_\circ,\rho_\bullet)= z_\alpha(z_\beta-1)+\rho_\circ(z_\alpha-z_\beta)\\
	\label{1Jrz}
	J_\bullet(\rho_\circ,\rho_\bullet)= z_\beta(1-z_\alpha)+\rho_\bullet(z_\alpha-z_\beta)
	\end{gather}
where $z_\alpha \in [0,\min(1,\alpha)]$ and $z_\beta \in [0,\min(1,\beta)]$ are solution of the saddle point equations
	\begin{gather}
	\frac{\rho_\circ}{z_\alpha}+\frac{\rho_\bullet}{z_\alpha-1}+\frac{1-\rho_\circ-\rho_\bullet}{z_\alpha-\alpha}=0\\ \label{chang-var2}
	\frac{\rho_\bullet}{z_\beta}+\frac{\rho_\circ}{z_\beta-1}+\frac{1-\rho_\circ-\rho_\bullet}{z_\beta-\beta}=0.
	\end{gather}
	The variables $z_{\alpha}, z_{\beta}$ happen to be the Riemann variables for this model \cite{cantini2022hydrodynamic}. 
In figure 	\ref{fig:2tasep} (left) we reported two 
examples of simulations of the $2$-TASEP on a lattice of 
size $L=100$ and with different values of the model 
parameters. We see that the numerical result agrees very 
well with the theoretical prediction obtained through the iterative solution of eqs.(\ref{principle},\ref{equ:curr equa}). The convergence of the iterative procedure is reported on the right of the 
same figure.

\subsubsection{Phase diagram}

\begin{figure}[h!]
\centering
		\includegraphics[width=.49\linewidth]{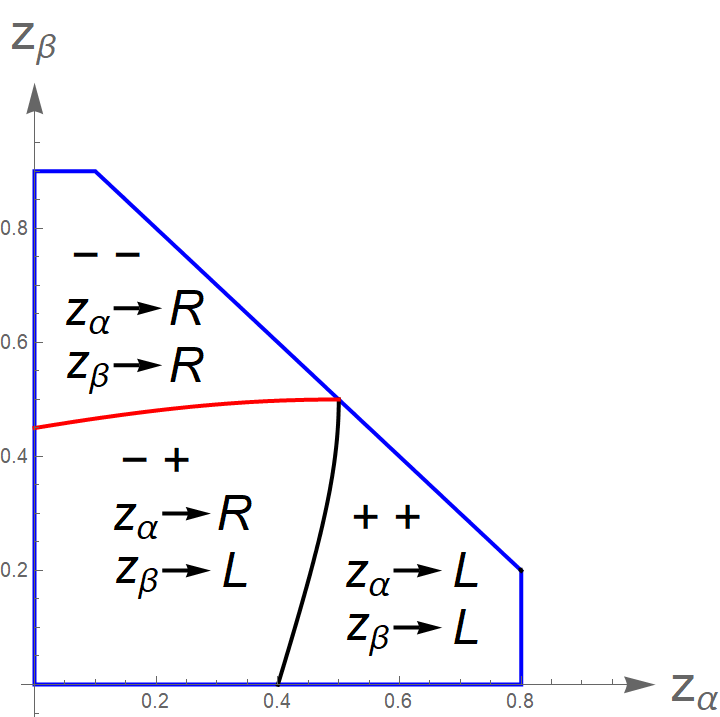}
		\includegraphics[width=.49\linewidth]{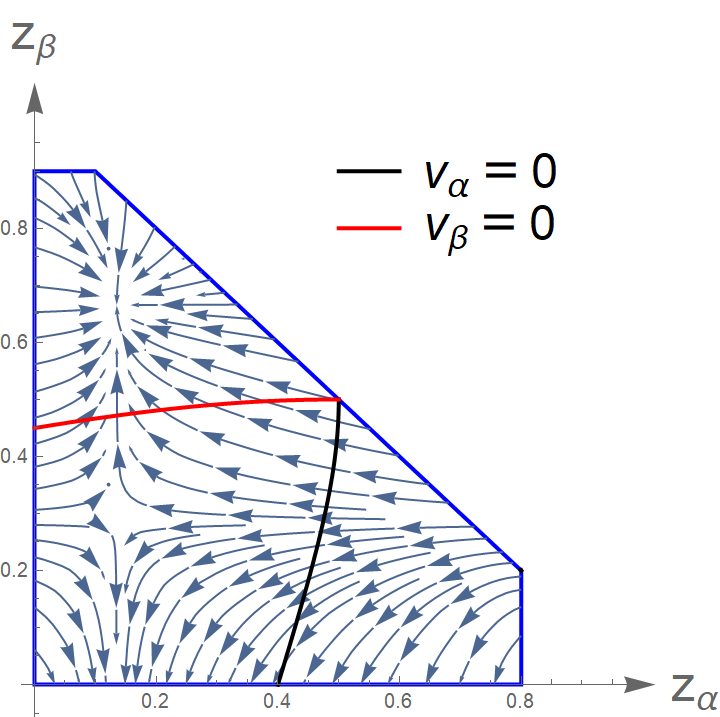}
	\caption{Phase diagram of a 2-species TASEP ($\alpha=0.8,\beta=0.9$). The signs on the left correspond to the velocities $v_\alpha$ and $v_\beta$ in order. On the right, we have an example of the ODE flow exhibiting a sink singularity in the left-induced phase and a saddle point in the mixed-induced phase.}
	\label{fig:phase}
\end{figure}

Following the discussion in Section \ref{sect:MCQ} we 
partition the phase space of the bulk densities of this
model in phases, characterized  by the sign of the functions $v_k(\mathbf{z}^B)$.
This a priori results in 9 phases for a two-component system, however, hyperbolicity of the corresponding conservation laws implies that some phases are forbidden as illustrated in the following table
\begin{center}
   \begin{tabular}{ c || c | c | c }
      & $v_{\alpha}<0$ & $v_{\alpha} = 0$ & $v_{\alpha}>0$ \\ \hline
      \hline
     $v_{\beta}<0$ & $R R$ & $B R$ & $L R$ \\ \hline
    $v_{\beta}=0$ & $\times$ & $B B$ & $L B$ \\ \hline
     $v_{\beta}>0$ & $\times$ & $\times$ & $L L$ \\
   \end{tabular}
\end{center}
In the preceding table the first letter represents the state of $z_{\alpha}$: L: left induced, R: right induced, B: bulk induced. The second letter is for the state of $z_{\beta}$. The symbol $\times$ is for a forbidden phase. See figure \ref{fig:phase} for the result of this partitioning  
for the values $\alpha=0.8,\beta=0.9$ of the bulk exchange rates.
%

\begin{figure}
\centering
\includegraphics[width=.49\linewidth]{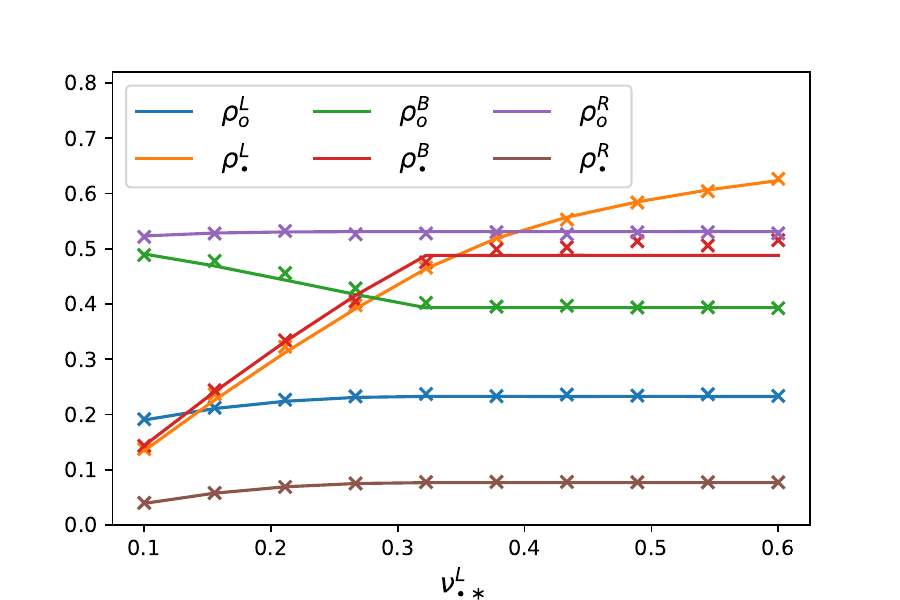}
\includegraphics[width=.49\linewidth]{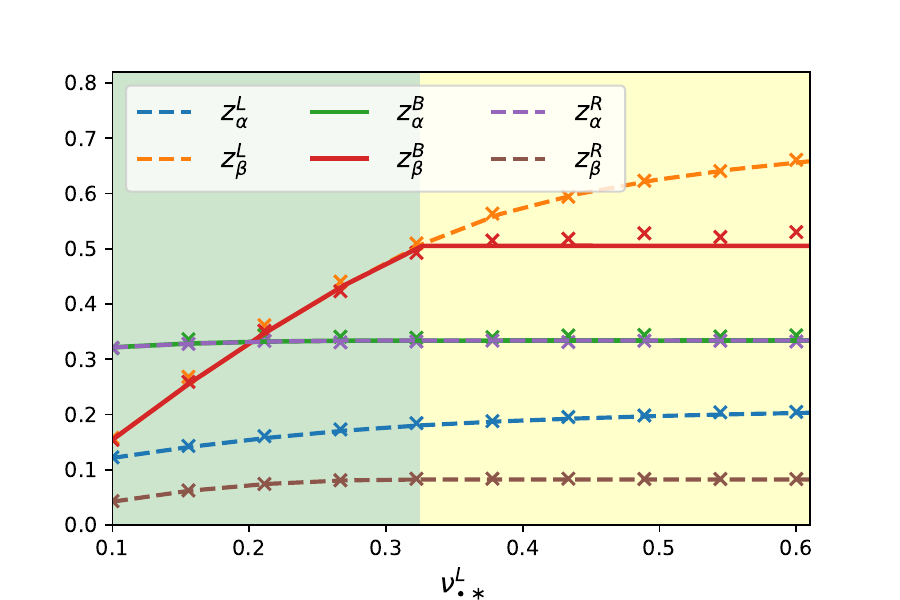}
\caption{Bulk and boundary densities (left) and the corresponding Riemann variables (right) of 2-TASEP as a function of the $\nu_{\bullet \ast}^{L}$. The crosses represent the numerical simulations, while the lines are the theoretical predictions. For the green shaded region $v_{\beta} >0$, while for the yellow shaded section $v_{\beta} = 0$ (in both regions $v_{\alpha} < 0$).}
\label{fig:2tasep cut}
\end{figure}

Numerical evidence for this diagram is reported in figure \ref{fig:2tasep cut}, where the results of simulations are shown together with theoretical predictions with varying parameter $\nu^L_{\bullet \ast}$ and all the other parameters fixed.
We notice that $z_{\beta}^{B}$ coincides with $z_{\beta}^{L}$ within the region where $v_{\beta}<0$, and they split in the region where $v_{\beta}=0$. At the same time $z_{\alpha}^L$ coincides with $z_{\alpha}^B$ for both regions since $v_{\alpha} < 0$.

\begin{figure}[h!]
\centering
\includegraphics[width=.6\linewidth]{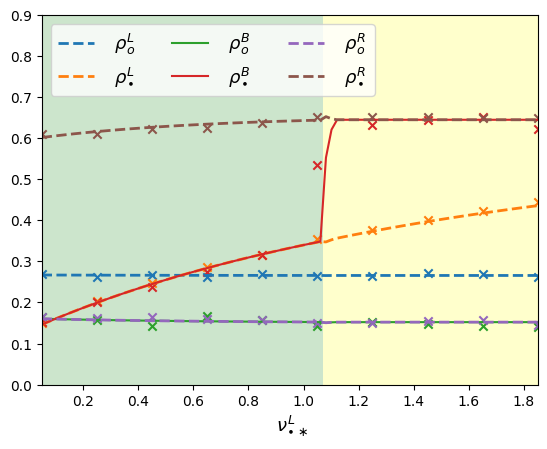}
	\caption{Bulk and boundary densities of the 2-ASEP as a function of the $\nu_{\bullet \ast}^{L}$. The crosses represent the numerical simulations, while the lines are the theoretical predictions. For the green shaded region $v_\bullet >0$, while for the yellow shaded section $v_\bullet < 0$ (in both regions $v_\circ < 0$). 
Parameters values: $q=0.5, \nu^L_{\ast\circ }=0.9,
\nu^L_{\bullet\circ }=\nu^L_{\ast\bullet}=1,
\nu^L_{\circ\ast}=\nu^L_{\circ\bullet  }=0.6, \nu^R_{\circ 
\ast}=\nu^R_{\bullet \circ}=0.1,\nu^R_{\circ \bullet}
=
\nu^R_{\ast\circ}=0.3,\nu^R_{\bullet\ast}=0.4,\nu^R_{ 
\ast\bullet}=0.8 $.
}%
	\label{fig:2asep}
\end{figure}

\subsection{2-species ASEP and 3-species TASEP}

We have considered other two models for which we have
access to the exact expressions of the hydrodynamic current as functions of the densities,.

\vspace{.3cm}

The first model, a $2$-species ASEP, contains two species of particles and the following bulk exchange rates:
\begin{equation}
\nu_{ij} =
\begin{cases}
1 & \text{if } \quad i > j\\
q & \text{if } \quad i < j
    \end{cases}
\end{equation}
where we have chosen the following order on the species: $\bullet > \ast > \circ$.

Although the stationary measure for a uniform state is not a product measure, yet, it's straightforward to write the currents-density relations since each of the $\bullet$ and $\circ$ particles dynamics can be decoupled in the bulk:
\begin{equation}
\begin{split}
J_{\bullet} = (1-q)\rho_{\bullet}(1-\rho_{\bullet}) \\
J_{\circ} = (q-1)\rho_{\circ}(1-\rho_{\circ}) .
\end{split}
\end{equation}
From these equations it is immediate that the densities are also Riemann variables for this model.
However, the dynamics of the two species cannot in general be decoupled on the boundaries, making the max-min principle not applicable in this case.

\vspace{.3cm}

The last model we have considered, a $3$-species TASEP, contains particles with labels $(1,2,3,4)$, where the type $4$ can be seen as empty sites, and bulk hopping rates:
\begin{equation}i j \xrightarrow[]{\nu_{ij}} ji\qquad
\nu_{ij} =
\begin{cases}
0 & \text{if } \quad i > j\\
\nu_{12} & \text{if } \quad (i,j) = (1,2) \\
\nu_{34} & \text{if } \quad (i,j) = (3,4) \\
1 & \text{otherwise}
    \end{cases}
\end{equation}

The particle currents of this model can be derived 
from those of the $2$-TASEP, $J_{\circ/\bullet}(\rho_\circ,
\rho_\bullet,\alpha,\beta)$, by making some particle 
identifications. Firstly, the particles $4$ and $3$ can be 
seen as $\circ$,  $1$ as $\bullet$ and $2$ as $\ast$, for $
\alpha = 1,\beta=\nu_{12}$. Secondly, $1$ and $2$ can be 
seen as $\bullet$, $3$ as $\ast$ and $4$ as $\circ$ with $
\alpha = \nu_{34},\beta=1$. Using densities of particles of species $1,2$ and $4$ as independent variables one finds
\begin{equation}
\begin{split}
J_1 & = J_{\bullet}(1-\rho_{1}-\rho_{2}, \rho_1,1,\nu_{12})\\
J_2 & = J_{\bullet}(\rho_4, \rho_1+\rho_2,\nu_{34},1)-J_1\\
J_4 & = J_{\circ}(\rho_{4},\rho_{1}+\rho_{2}, \nu_{34},1). 
\end{split}
\end{equation}

\begin{figure}[h]
\centering
\includegraphics[width=.6\linewidth]{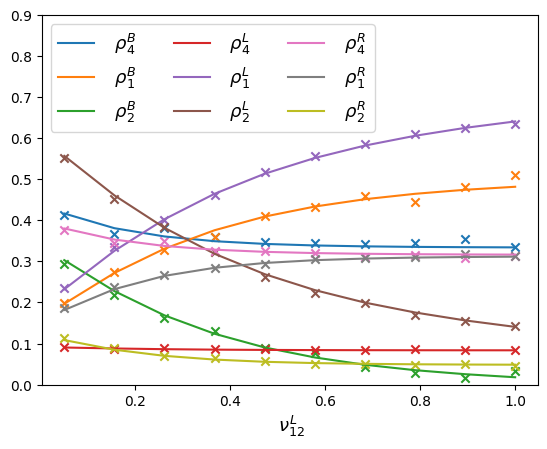}
	\caption{Bulk and boundary densities of 3-TASEP as a function of the parameter $\nu_{12}^{L}$. The crosses represent the numerical simulations, while the continuous lines are the theoretical predictions.  
Parameters values: $\nu_{12}=0.6,\nu_{34}=1.3,
\nu^L_{13}=0.3,
\nu^L_{14}= 0.9,\nu^L_{23}=0.5,
\nu^L_{24}= 0.8,\nu^L_{34}=1, 
\nu^R_{12}=0.2,\nu^R_{13}=0.4,\nu^R_{14}=0.2,
\nu^R_{23}=0.6,\nu^R_{24}=0.7,\nu^R_{ 
34}=0.4 $. }
	\label{fig:3tasep}
\end{figure}

In figure \ref{fig:2asep} and \ref{fig:3tasep} we report the results for the bulk and boundary densities of these models, obtained though simulations of a system of size $L=100$, along with the theoretical predictions. One boundary parameter is varied ($\nu^L_{\bullet \ast}$ in the $2$-ASEP and $\nu^L_{12}$ in the $3$-TASEP) while all the other parameters are fixed.
Similarly to the case of the $2$--TASEP seen in the previous section, we find good agreement.

\subsection{Conclusion}

In conclusion, this paper introduces a method which allows to determine the steady state average particle densities and currents of one-dimensional multi-species driven system with open boundaries.
The method, rooted in the bulk hydrodynamic behavior of the model, extends the max-min principle applicable to single-species models \cite{krug1991boundary,
krug1991steady,popkov1999steady,hager2001minimal}.
By comparing our method's predictions with numerical simulations across three models, we observed good agreement.
Our analysis of bulk hydrodynamic conservation laws enables us to predict the phase diagram, which becomes more intelligible when considering the behavior of the Riemann variables of the model (when they exist).

The major open question  pertains to the method's domain of validity, particularly in establishing precise definitions of boundary densities for more general boundary conditions. 
The heuristic argument in favor  of our method rests on the existence of a complete set of Riemann variables in bulk dynamics. Therefore, exploring models with more than two species, lacking this completeness, and subjecting our method to such models, presents an intriguing avenue for future research.

\section*{ACKNOWLEDGMENTS}

We thank Gunter Sch\"utz for useful discussions.
The work of A. Zhara has been partially funded by the ERC Starting Grant 101042293 (HEPIQ) and completed while he was a member of LPTM.

\bibliography{f}

\begin{thebibliography}{22}%
\makeatletter
\providecommand \@ifxundefined [1]{%
 \@ifx{#1\undefined}
}%
\providecommand \@ifnum [1]{%
 \ifnum #1\expandafter \@firstoftwo
 \else \expandafter \@secondoftwo
 \fi
}%
\providecommand \@ifx [1]{%
 \ifx #1\expandafter \@firstoftwo
 \else \expandafter \@secondoftwo
 \fi
}%
\providecommand \natexlab [1]{#1}%
\providecommand \enquote  [1]{``#1''}%
\providecommand \bibnamefont  [1]{#1}%
\providecommand \bibfnamefont [1]{#1}%
\providecommand \citenamefont [1]{#1}%
\providecommand \href@noop [0]{\@secondoftwo}%
\providecommand \href [0]{\begingroup \@sanitize@url \@href}%
\providecommand \@href[1]{\@@startlink{#1}\@@href}%
\providecommand \@@href[1]{\endgroup#1\@@endlink}%
\providecommand \@sanitize@url [0]{\catcode `\\12\catcode `\$12\catcode
  `\&12\catcode `\#12\catcode `\^12\catcode `\_12\catcode `\%12\relax}%
\providecommand \@@startlink[1]{}%
\providecommand \@@endlink[0]{}%
\providecommand \url  [0]{\begingroup\@sanitize@url \@url }%
\providecommand \@url [1]{\endgroup\@href {#1}{\urlprefix }}%
\providecommand \urlprefix  [0]{URL }%
\providecommand \Eprint [0]{\href }%
\providecommand \doibase [0]{https://doi.org/}%
\providecommand \selectlanguage [0]{\@gobble}%
\providecommand \bibinfo  [0]{\@secondoftwo}%
\providecommand \bibfield  [0]{\@secondoftwo}%
\providecommand \translation [1]{[#1]}%
\providecommand \BibitemOpen [0]{}%
\providecommand \bibitemStop [0]{}%
\providecommand \bibitemNoStop [0]{.\EOS\space}%
\providecommand \EOS [0]{\spacefactor3000\relax}%
\providecommand \BibitemShut  [1]{\csname bibitem#1\endcsname}%
\let\auto@bib@innerbib\@empty
\bibitem [{\citenamefont {Chou}\ \emph {et~al.}(2011)\citenamefont {Chou},
  \citenamefont {Mallick},\ and\ \citenamefont {Zia}}]{chou2011non}%
  \BibitemOpen
  \bibfield  {author} {\bibinfo {author} {\bibfnamefont {T.}~\bibnamefont
  {Chou}}, \bibinfo {author} {\bibfnamefont {K.}~\bibnamefont {Mallick}},\ and\
  \bibinfo {author} {\bibfnamefont {R.~K.~P.}\ \bibnamefont {Zia}},\ }\bibfield
   {title} {\bibinfo {title} {Non-equilibrium statistical mechanics: from a
  paradigmatic model to biological transport},\ }\href@noop {} {\bibfield
  {journal} {\bibinfo  {journal} {Reports on progress in physics}\ }\textbf
  {\bibinfo {volume} {74}},\ \bibinfo {pages} {116601} (\bibinfo {year}
  {2011})}\BibitemShut {NoStop}%
\bibitem [{\citenamefont {Blythe}\ and\ \citenamefont
  {Evans}(2007)}]{blythe2007nonequilibrium}%
  \BibitemOpen
  \bibfield  {author} {\bibinfo {author} {\bibfnamefont {R.~A.}\ \bibnamefont
  {Blythe}}\ and\ \bibinfo {author} {\bibfnamefont {M.~R.}\ \bibnamefont
  {Evans}},\ }\bibfield  {title} {\bibinfo {title} {Nonequilibrium steady
  states of matrix-product form: a solver's guide},\ }\href@noop {} {\bibfield
  {journal} {\bibinfo  {journal} {Journal of Physics A: Mathematical and
  Theoretical}\ }\textbf {\bibinfo {volume} {40}},\ \bibinfo {pages} {R333}
  (\bibinfo {year} {2007})}\BibitemShut {NoStop}%
\bibitem [{\citenamefont {Fang}\ \emph {et~al.}(2019)\citenamefont {Fang},
  \citenamefont {Kruse}, \citenamefont {Lu},\ and\ \citenamefont
  {Wang}}]{fang2019nonequilibrium}%
  \BibitemOpen
  \bibfield  {author} {\bibinfo {author} {\bibfnamefont {X.}~\bibnamefont
  {Fang}}, \bibinfo {author} {\bibfnamefont {K.}~\bibnamefont {Kruse}},
  \bibinfo {author} {\bibfnamefont {T.}~\bibnamefont {Lu}},\ and\ \bibinfo
  {author} {\bibfnamefont {J.}~\bibnamefont {Wang}},\ }\bibfield  {title}
  {\bibinfo {title} {Nonequilibrium physics in biology},\ }\href@noop {}
  {\bibfield  {journal} {\bibinfo  {journal} {Reviews of Modern Physics}\
  }\textbf {\bibinfo {volume} {91}},\ \bibinfo {pages} {045004} (\bibinfo
  {year} {2019})}\BibitemShut {NoStop}%
\bibitem [{\citenamefont {Schmittmann}\ and\ \citenamefont
  {Zia}(1995)}]{schmittmann1995statistical}%
  \BibitemOpen
  \bibfield  {author} {\bibinfo {author} {\bibfnamefont {B.}~\bibnamefont
  {Schmittmann}}\ and\ \bibinfo {author} {\bibfnamefont {R.~K.-P.}\
  \bibnamefont {Zia}},\ }\bibfield  {title} {\bibinfo {title} {Statistical
  mechanics of driven diffusive systems},\ }\href@noop {} {\bibfield  {journal}
  {\bibinfo  {journal} {Phase transitions and critical phenomena}\ }\textbf
  {\bibinfo {volume} {17}},\ \bibinfo {pages} {3} (\bibinfo {year}
  {1995})}\BibitemShut {NoStop}%
\bibitem [{\citenamefont {Krug}(1991{\natexlab{a}})}]{krug1991boundary}%
  \BibitemOpen
  \bibfield  {author} {\bibinfo {author} {\bibfnamefont {J.}~\bibnamefont
  {Krug}},\ }\bibfield  {title} {\bibinfo {title} {Boundary--induced phase
  transitions in driven diffusive systems},\ }\href@noop {} {\bibfield
  {journal} {\bibinfo  {journal} {Physical review letters}\ }\textbf {\bibinfo
  {volume} {67}},\ \bibinfo {pages} {1882} (\bibinfo {year}
  {1991}{\natexlab{a}})}\BibitemShut {NoStop}%
\bibitem [{\citenamefont {Krug}(1991{\natexlab{b}})}]{krug1991steady}%
  \BibitemOpen
  \bibfield  {author} {\bibinfo {author} {\bibfnamefont {J.}~\bibnamefont
  {Krug}},\ }\bibfield  {title} {\bibinfo {title} {Steady state selection in
  driven diffusive systems},\ }in\ \href@noop {} {\emph {\bibinfo {booktitle}
  {Spontaneous formation of space-time structures and criticality}}}\ (\bibinfo
   {publisher} {Springer},\ \bibinfo {year} {1991})\ pp.\ \bibinfo {pages}
  {37--40}\BibitemShut {NoStop}%
\bibitem [{\citenamefont {Popkov}\ and\ \citenamefont
  {Sch{\"u}tz}(1999)}]{popkov1999steady}%
  \BibitemOpen
  \bibfield  {author} {\bibinfo {author} {\bibfnamefont {V.}~\bibnamefont
  {Popkov}}\ and\ \bibinfo {author} {\bibfnamefont {G.~M.}\ \bibnamefont
  {Sch{\"u}tz}},\ }\bibfield  {title} {\bibinfo {title} {Steady-state selection
  in driven diffusive systems with open boundaries},\ }\href@noop {} {\bibfield
   {journal} {\bibinfo  {journal} {EPL (Europhysics Letters)}\ }\textbf
  {\bibinfo {volume} {48}},\ \bibinfo {pages} {257} (\bibinfo {year}
  {1999})}\BibitemShut {NoStop}%
\bibitem [{\citenamefont {Hager}\ \emph {et~al.}(2001)\citenamefont {Hager},
  \citenamefont {Krug}, \citenamefont {Popkov},\ and\ \citenamefont
  {Sch{\"u}tz}}]{hager2001minimal}%
  \BibitemOpen
  \bibfield  {author} {\bibinfo {author} {\bibfnamefont {J.}~\bibnamefont
  {Hager}}, \bibinfo {author} {\bibfnamefont {J.}~\bibnamefont {Krug}},
  \bibinfo {author} {\bibfnamefont {V.}~\bibnamefont {Popkov}},\ and\ \bibinfo
  {author} {\bibfnamefont {G.}~\bibnamefont {Sch{\"u}tz}},\ }\bibfield  {title}
  {\bibinfo {title} {Minimal current phase and universal boundary layers in
  driven diffusive systems},\ }\href@noop {} {\bibfield  {journal} {\bibinfo
  {journal} {Physical Review E}\ }\textbf {\bibinfo {volume} {63}},\ \bibinfo
  {pages} {056110} (\bibinfo {year} {2001})}\BibitemShut {NoStop}%
\bibitem [{\citenamefont {R{\'a}kos}\ and\ \citenamefont
  {Sch{\"u}tz}(2004)}]{rakos2004exact}%
  \BibitemOpen
  \bibfield  {author} {\bibinfo {author} {\bibfnamefont {A.}~\bibnamefont
  {R{\'a}kos}}\ and\ \bibinfo {author} {\bibfnamefont {G.}~\bibnamefont
  {Sch{\"u}tz}},\ }\bibfield  {title} {\bibinfo {title} {Exact shock measures
  and steady-state selection in a driven diffusive system with two conserved
  densities},\ }\href@noop {} {\bibfield  {journal} {\bibinfo  {journal}
  {Journal of statistical physics}\ }\textbf {\bibinfo {volume} {117}},\
  \bibinfo {pages} {55} (\bibinfo {year} {2004})}\BibitemShut {NoStop}%
\bibitem [{\citenamefont {Popkov}(2004)}]{popkov2004infinite}%
  \BibitemOpen
  \bibfield  {author} {\bibinfo {author} {\bibfnamefont {V.}~\bibnamefont
  {Popkov}},\ }\bibfield  {title} {\bibinfo {title} {Infinite reflections of
  shock fronts in driven diffusive systems with two species},\ }\href@noop {}
  {\bibfield  {journal} {\bibinfo  {journal} {Journal of Physics A:
  Mathematical and General}\ }\textbf {\bibinfo {volume} {37}},\ \bibinfo
  {pages} {1545} (\bibinfo {year} {2004})}\BibitemShut {NoStop}%
\bibitem [{\citenamefont {Bonnin}\ \emph {et~al.}(2021)\citenamefont {Bonnin},
  \citenamefont {Stansfield}, \citenamefont {Romano},\ and\ \citenamefont
  {Kern}}]{bonnin2021two}%
  \BibitemOpen
  \bibfield  {author} {\bibinfo {author} {\bibfnamefont {P.}~\bibnamefont
  {Bonnin}}, \bibinfo {author} {\bibfnamefont {I.}~\bibnamefont {Stansfield}},
  \bibinfo {author} {\bibfnamefont {M.~C.}\ \bibnamefont {Romano}},\ and\
  \bibinfo {author} {\bibfnamefont {N.}~\bibnamefont {Kern}},\ }\bibfield
  {title} {\bibinfo {title} {Two-species tasep model: from a simple description
  to intermittency and travelling traffic jams},\ }\href@noop {} {\bibfield
  {journal} {\bibinfo  {journal} {arXiv preprint arXiv:2102.02486}\ } (\bibinfo
  {year} {2021})}\BibitemShut {NoStop}%
\bibitem [{\citenamefont {Gupta}\ \emph {et~al.}(2023)\citenamefont {Gupta},
  \citenamefont {Pal},\ and\ \citenamefont {Gupta}}]{gupta2023interplay}%
  \BibitemOpen
  \bibfield  {author} {\bibinfo {author} {\bibfnamefont {A.}~\bibnamefont
  {Gupta}}, \bibinfo {author} {\bibfnamefont {B.}~\bibnamefont {Pal}},\ and\
  \bibinfo {author} {\bibfnamefont {A.~K.}\ \bibnamefont {Gupta}},\ }\bibfield
  {title} {\bibinfo {title} {Interplay of reservoirs in a bidirectional
  system},\ }\href@noop {} {\bibfield  {journal} {\bibinfo  {journal} {Physical
  Review E}\ }\textbf {\bibinfo {volume} {107}},\ \bibinfo {pages} {034103}
  (\bibinfo {year} {2023})}\BibitemShut {NoStop}%
\bibitem [{\citenamefont {Bardos}\ \emph {et~al.}(1979)\citenamefont {Bardos},
  \citenamefont {LeRoux},\ and\ \citenamefont
  {N{\'e}d{\'e}lec}}]{bardos1979first}%
  \BibitemOpen
  \bibfield  {author} {\bibinfo {author} {\bibfnamefont {C.}~\bibnamefont
  {Bardos}}, \bibinfo {author} {\bibfnamefont {A.-Y.}\ \bibnamefont {LeRoux}},\
  and\ \bibinfo {author} {\bibfnamefont {J.-C.}\ \bibnamefont
  {N{\'e}d{\'e}lec}},\ }\bibfield  {title} {\bibinfo {title} {First order
  quasilinear equations with boundary conditions},\ }\href@noop {} {\bibfield
  {journal} {\bibinfo  {journal} {Communications in partial differential
  equations}\ }\textbf {\bibinfo {volume} {4}},\ \bibinfo {pages} {1017}
  (\bibinfo {year} {1979})}\BibitemShut {NoStop}%
\bibitem [{\citenamefont {Dubois}\ and\ \citenamefont
  {Le~Floch}(1988)}]{dubois1988boundary}%
  \BibitemOpen
  \bibfield  {author} {\bibinfo {author} {\bibfnamefont {F.}~\bibnamefont
  {Dubois}}\ and\ \bibinfo {author} {\bibfnamefont {P.}~\bibnamefont
  {Le~Floch}},\ }\bibfield  {title} {\bibinfo {title} {Boundary conditions for
  nonlinear hyperbolic systems of conservation laws},\ }\href@noop {}
  {\bibfield  {journal} {\bibinfo  {journal} {Journal of Differential
  Equations}\ }\textbf {\bibinfo {volume} {71}},\ \bibinfo {pages} {93}
  (\bibinfo {year} {1988})}\BibitemShut {NoStop}%
\bibitem [{\citenamefont {Mazet}\ and\ \citenamefont
  {Bourdel}(1986)}]{mazet1986analyse}%
  \BibitemOpen
  \bibfield  {author} {\bibinfo {author} {\bibfnamefont {P.}~\bibnamefont
  {Mazet}}\ and\ \bibinfo {author} {\bibfnamefont {F.}~\bibnamefont
  {Bourdel}},\ }\bibfield  {title} {\bibinfo {title} {Analyse num{\'e}rique des
  {\'e}quations d'euler pour l'{\'e}tude des {\'e}coulements autour de corps
  {\'e}lanc{\'e}s en incidence},\ }\href@noop {} {\bibfield  {journal}
  {\bibinfo  {journal} {CERT Report}\ } (\bibinfo {year} {1986})}\BibitemShut
  {NoStop}%
\bibitem [{\citenamefont {Katz}\ \emph {et~al.}()\citenamefont {Katz},
  \citenamefont {Lebowitz},\ and\ \citenamefont
  {Spohn}}]{katz_nonequilibrium_1984}%
  \BibitemOpen
  \bibfield  {author} {\bibinfo {author} {\bibfnamefont {S.}~\bibnamefont
  {Katz}}, \bibinfo {author} {\bibfnamefont {J.~L.}\ \bibnamefont {Lebowitz}},\
  and\ \bibinfo {author} {\bibfnamefont {H.}~\bibnamefont {Spohn}},\ }\bibfield
   {title} {\bibinfo {title} {Nonequilibrium steady states of stochastic
  lattice gas models of fast ionic conductors},\ }\href
  {https://doi.org/10.1007/BF01018556} {\ \textbf {\bibinfo {volume} {34}},\
  \bibinfo {pages} {497}}\BibitemShut {NoStop}%
\bibitem [{\citenamefont {Osher}(1983)}]{osher1983riemann}%
  \BibitemOpen
  \bibfield  {author} {\bibinfo {author} {\bibfnamefont {S.}~\bibnamefont
  {Osher}},\ }\bibfield  {title} {\bibinfo {title} {The riemann problem for
  nonconvex scalar conservation laws and hamilton-jacobi equations},\
  }\href@noop {} {\bibfield  {journal} {\bibinfo  {journal} {Proceedings of the
  American Mathematical Society}\ }\textbf {\bibinfo {volume} {89}},\ \bibinfo
  {pages} {641} (\bibinfo {year} {1983})}\BibitemShut {NoStop}%
\bibitem [{\citenamefont {Popkov}\ and\ \citenamefont
  {Salerno}(2011)}]{popkov2011hierarchy}%
  \BibitemOpen
  \bibfield  {author} {\bibinfo {author} {\bibfnamefont {V.}~\bibnamefont
  {Popkov}}\ and\ \bibinfo {author} {\bibfnamefont {M.}~\bibnamefont
  {Salerno}},\ }\bibfield  {title} {\bibinfo {title} {Hierarchy of
  boundary-driven phase transitions in multispecies particle systems},\
  }\href@noop {} {\bibfield  {journal} {\bibinfo  {journal} {Physical Review
  E}\ }\textbf {\bibinfo {volume} {83}},\ \bibinfo {pages} {011130} (\bibinfo
  {year} {2011})}\BibitemShut {NoStop}%
\bibitem [{\citenamefont {Derrida}(1996)}]{derrida1996statphys}%
  \BibitemOpen
  \bibfield  {author} {\bibinfo {author} {\bibfnamefont {B.}~\bibnamefont
  {Derrida}},\ }\bibfield  {title} {\bibinfo {title} {{Statphys-19: 19th IUPAP
  Int}},\ }\bibfield  {booktitle} {\emph {\bibinfo {booktitle} {Conf. on
  Statistical Physics (Xiamen 1996) ed BL Hao (Singapore: World Scientific)}},\
  }\href@noop {} {\  (\bibinfo {year} {1996})}\BibitemShut {NoStop}%
\bibitem [{\citenamefont {Mallick}(1996)}]{mallick1996shocks}%
  \BibitemOpen
  \bibfield  {author} {\bibinfo {author} {\bibfnamefont {K.}~\bibnamefont
  {Mallick}},\ }\bibfield  {title} {\bibinfo {title} {Shocks in the asymmetry
  exclusion model with an impurity},\ }\href@noop {} {\bibfield  {journal}
  {\bibinfo  {journal} {Journal of Physics A: Mathematical and General}\
  }\textbf {\bibinfo {volume} {29}},\ \bibinfo {pages} {5375} (\bibinfo {year}
  {1996})}\BibitemShut {NoStop}%
\bibitem [{\citenamefont {Cantini}(2008)}]{cantini2008algebraic}%
  \BibitemOpen
  \bibfield  {author} {\bibinfo {author} {\bibfnamefont {L.}~\bibnamefont
  {Cantini}},\ }\bibfield  {title} {\bibinfo {title} {Algebraic bethe ansatz
  for the two species asep with different hopping rates},\ }\href@noop {}
  {\bibfield  {journal} {\bibinfo  {journal} {Journal of Physics A:
  Mathematical and Theoretical}\ }\textbf {\bibinfo {volume} {41}},\ \bibinfo
  {pages} {095001} (\bibinfo {year} {2008})}\BibitemShut {NoStop}%
\bibitem [{\citenamefont {Cantini}\ and\ \citenamefont
  {Zahra}(2022)}]{cantini2022hydrodynamic}%
  \BibitemOpen
  \bibfield  {author} {\bibinfo {author} {\bibfnamefont {L.}~\bibnamefont
  {Cantini}}\ and\ \bibinfo {author} {\bibfnamefont {A.}~\bibnamefont
  {Zahra}},\ }\bibfield  {title} {\bibinfo {title} {Hydrodynamic behavior of
  the two-tasep},\ }\href@noop {} {\bibfield  {journal} {\bibinfo  {journal}
  {Journal of Physics A: Mathematical and Theoretical}\ }\textbf {\bibinfo
  {volume} {55}},\ \bibinfo {pages} {305201} (\bibinfo {year}
  {2022})}\BibitemShut {NoStop}%
\end{thebibliography}%
\end{document}